\documentclass[epj]{svjour}
\usepackage{graphics}
\begin{document}
\title{Parameter-independent predictions for shape variables of heavy deformed nuclei 
in the proxy-SU(3) model}

\author{Dennis Bonatsos\inst{1} \and I. E. Assimakis \inst{1} \and N. Minkov\inst{2} \and \underline{Andriana Martinou}\inst{1} \and S. Sarantopoulou\inst{1} \and R. B. Cakirli\inst{3} \and R. F. Casten\inst{4,5} \and
K. Blaum\inst{6}} 
%
%
\institute{Institute of Nuclear and Particle Physics, National Centre for Scientific Research 
``Demokritos'', GR-15310 Aghia Paraskevi, Attiki, Greece
\and Institute of Nuclear Research and Nuclear Energy, Bulgarian Academy of Sciences, 72 Tzarigrad Road, 1784 Sofia, Bulgaria
\and Department of Physics, University of Istanbul, Istanbul, Turkey
\and Wright Laboratory, Yale University, New Haven, Connecticut 06520, USA
\and Facility for Rare Isotope Beams, 640 South Shaw Lane, Michigan State University, East Lansing, MI 48824 USA
\and  Max-Planck-Institut f\"{u}r Kernphysik, Saupfercheckweg 1, D-69117 Heidelberg, Germany}

\date{Received: date / Revised version: date}
%
\abstract{
Using a new approximate analytic parameter-free proxy-SU(3) scheme, we make predictions of shape observables for deformed nuclei, namely $\beta$ and $\gamma$ deformation variables, 
and compare these with empirical data and with predictions by relativistic and non-relativistic mean-field theories. 
} 

\authorrunning{Dennis Bonatsos et al.}
\titlerunning{Shape variables in the proxy-SU(3) model}
\maketitle
\section{Introduction}
\label{intro}
Proxy-SU(3) is a new approximate symmetry scheme applicable in medium-mass and heavy deformed nuclei
\cite{PRC1,PRC2}. The basic features and the theoretical foundations of proxy-SU(3) have been described in Refs.  \cite{IoBonat,IoAssim}, to which the reader is referred. In this contribution we are going to focus attention on the first applications of proxy-SU(3) in making predictions for the deformation  variables of deformed rare earth nuclei. 

\section{Connection between deformation variables and SU(3) quantum numbers}

A connection between the collective variables $\beta$ and $\gamma$ of the collective model \cite{BM}
and the quantum numbers $\lambda$ and $\mu$ characterizing the irreducible represention 
$(\lambda,\mu)$ of SU(3) \cite{Elliott1,Elliott2} has long been established \cite{Castanos,Park}, based on the fact that the invariant quantities of the two theories should posses the same values. 

The relevant equation for $\beta$ reads 
\cite{Castanos,Park}
\begin{equation}\label{b1}
	\beta^2= {4\pi \over 5} {1\over (A \bar{r^2})^2} (\lambda^2+\lambda \mu + \mu^2+ 3\lambda +3 \mu +3), 
\end{equation}
where $A$ is the mass number of the nucleus and $\bar{r^2}$ is related to the dimensionless mean square radius \cite{Ring}, $\sqrt{\bar{r^2}}= r_0 A^{1/6}$. 
The constant $r_0$ is determined from a fit over a wide range of nuclei \cite{DeVries,Stone}. We use the value in Ref. \cite{Castanos}, $r_0=0.87$, in agreement to Ref. \cite{Stone}.
The quantity in Eq. (1) is proportional to the second order Casimir operator of SU(3) \cite{IA}, 
 \begin{equation}\label{C2} 
 C_2(\lambda,\mu)= {2 \over 3} (\lambda^2+\lambda \mu + \mu^2+ 3\lambda +3 \mu). 
\end{equation}

The relevant equation for $\gamma$ reads \cite{Castanos,Park}
\begin{equation}\label{g1}
\gamma = \arctan \left( {\sqrt{3} (\mu+1) \over 2\lambda+\mu+3}  \right). 
\end{equation}

\section{Predictions for the $\beta$ variable}

The $\beta$ deformation variable for a given nucleus can be obtained from Eq. (1), 
using the $(\lambda,\mu)$ values corresponding to the ground state band of this nucleus,
obtained from Table 2 of Ref. \cite{IoSarant}. 

A rescaling in order to take into account the size of the shell will be needed, as 
in the case of the geometric limit \cite{GK} of the Interacting Boson Model \cite{IA} 
in which a rescaling factor $2N_B/A$ is used, where $N_B$ is the number of bosons 
(half of the number of the valence nucleons measured from the closest closed shell)
in a nucleus with mass number $A$. In the present case one can see \cite{PRC2} that 
the $\beta$ values obtained from Eq. (1) should be multiplied by a rescaling factor 
$A/(S_p+S_n)$, where $S_p$ ($S_n$) is the size of the proton (neutron) shell in which the valence 
protons (neutrons) of the nucleus live. In the case of the rare earths considered here,
one has $S_p=82-50=32$ and $S_n=126-82=44$, thus the rescaling factor is $A/76$. 

Results for the $\beta$ variable for several isotopic chains are shown in Fig.~1. These can be compared to Relativistic Mean Field predictions \cite{Lalazissis} shown in Fig.~2, as well as to empirical 
$\beta$ values obtained from $B(E2)$ transition rates \cite{Raman} shown in Fig.~3. Indeed such detailed 
comparisons for various series of isotopes are shown in Figs.~4-7. We remark that the proxy-SU(3) 
predictions are in general in very good agreement with both the RMF predictions and the empirical values. The sudden minimum developed in Fig.~1 at $N=116$ could be related to the prolate-to-oblate 
shape/phase transition to be discussed in Ref.  \cite{IoSarant}. 

\section{Predictions for the $\gamma$ variable}

The $\gamma$ deformation variable for a given nucleus can be obtained from Eq. (3), 
using the $(\lambda,\mu)$ values corresponding to the ground state band of this nucleus,
obtained from Table 2 of Ref. \cite{IoSarant}. 

Results for the $\gamma$ variable for several isotopic chains are shown in Figs.~8 and 9. In Fig.~9,
predictions by  Gogny D1S calculations \cite{Robledo} are also shown for comparison. The sharp jump 
of the $\gamma$ variable from values close to 0 to values close to 60 degrees, seen in Fig.~9 
close to $N=116$, 
for both the proxy-SU(3) and the Gogny D1S predictions, can be related to the prolate-to-oblate 
shape/phase transition to be discussed in the next talk \cite{IoSarant}. In contrast, in the series 
of isotopes shown in Fig.~8, $\gamma$ is only raising at large neutron number $N$ up to 30 degrees,
indicating possible regions with triaxial shapes. 

Minima appear in the proxy-SU(3) predictions for the neutron numbers for which the relevant SU(3) irrep, seen in Table 2 of Ref. \cite{IoSarant}, happens to possess $\mu=0$, as one can easily see
from Eq. (3). These oscillations could probably be smoothed out through a procedure of taking 
the average of neighboring SU(3) representations, as in Ref. \cite{Bhatt}. 

Empirical values for the $\gamma$ variable can be estimated from ratios of the $\gamma$ vibrational bandhead to the first $2^+$ state,  
\begin{equation}
R={E(2^+_2)\over E(2^+_1)}, 
\end{equation}
through \cite{DF,Casten,Esser}
\begin{equation}
\sin 3\gamma= {3\over 2\sqrt{2}} \sqrt{1-\left({R-1\over R+1}  \right)^2}. 
\end{equation}
The proxy-SU(3) predictions for several isotopic chains are compared to so-obtained empirical values,
as well as to Gogny D1S predictions where available, in Figs.~10 and 11. Again in general good agreement is seen. 

\section{Conclusions} 

The proxy-SU(3) symmetry provides predictions for the $\beta$ collective variable which are in good agreement with RMF predictions, as well as with empirical values obtained from $B(E2)$ transition rates. Furthermore, the proxy-SU(3) symmetry provides predictions for the $\gamma$ collective variable which are in good agreement with Gogny D1S predictions, as well as with empirical values obtained from the $\gamma$ vibrational bandhead.

\begin{figure*}[htb]

\resizebox{0.99\textwidth}{!}{%

{\includegraphics{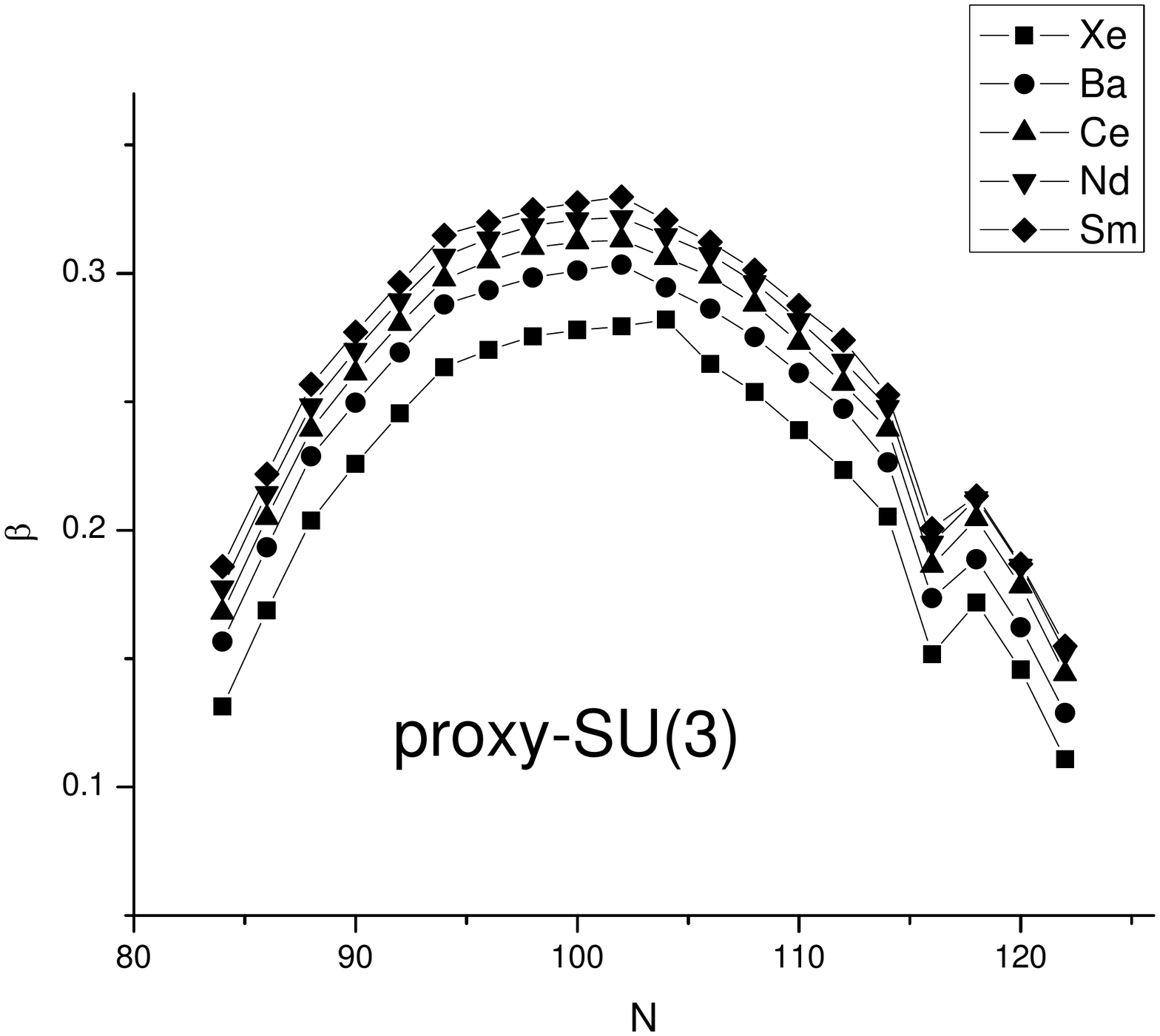}\hspace{5mm}
\includegraphics{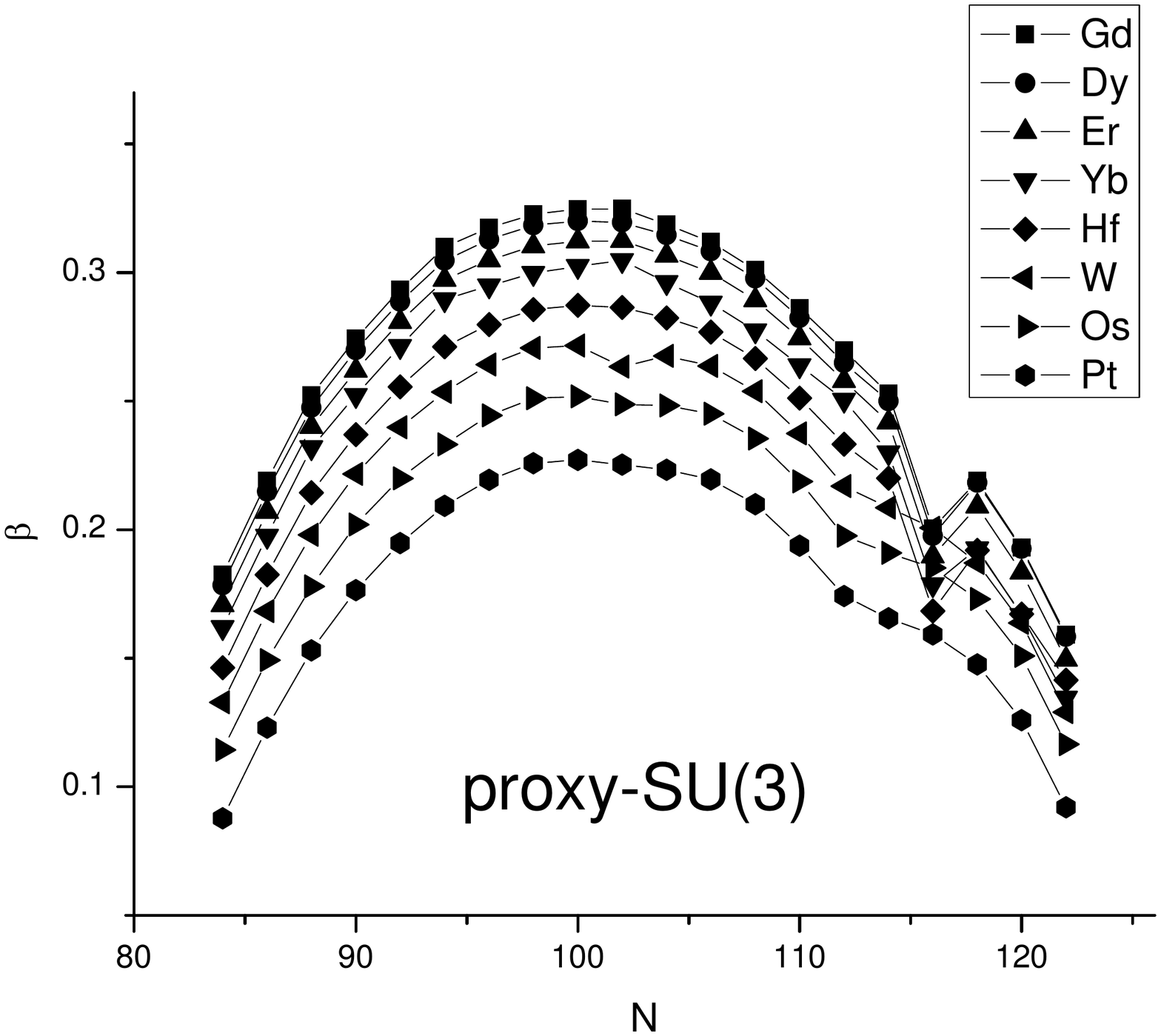}} 
}

\caption{Proxy-SU(3) predictions for $\beta$, obtained from Eq. (1).}
\end{figure*}

\begin{figure*}[htb]

\resizebox{0.99\textwidth}{!}{%

{\includegraphics{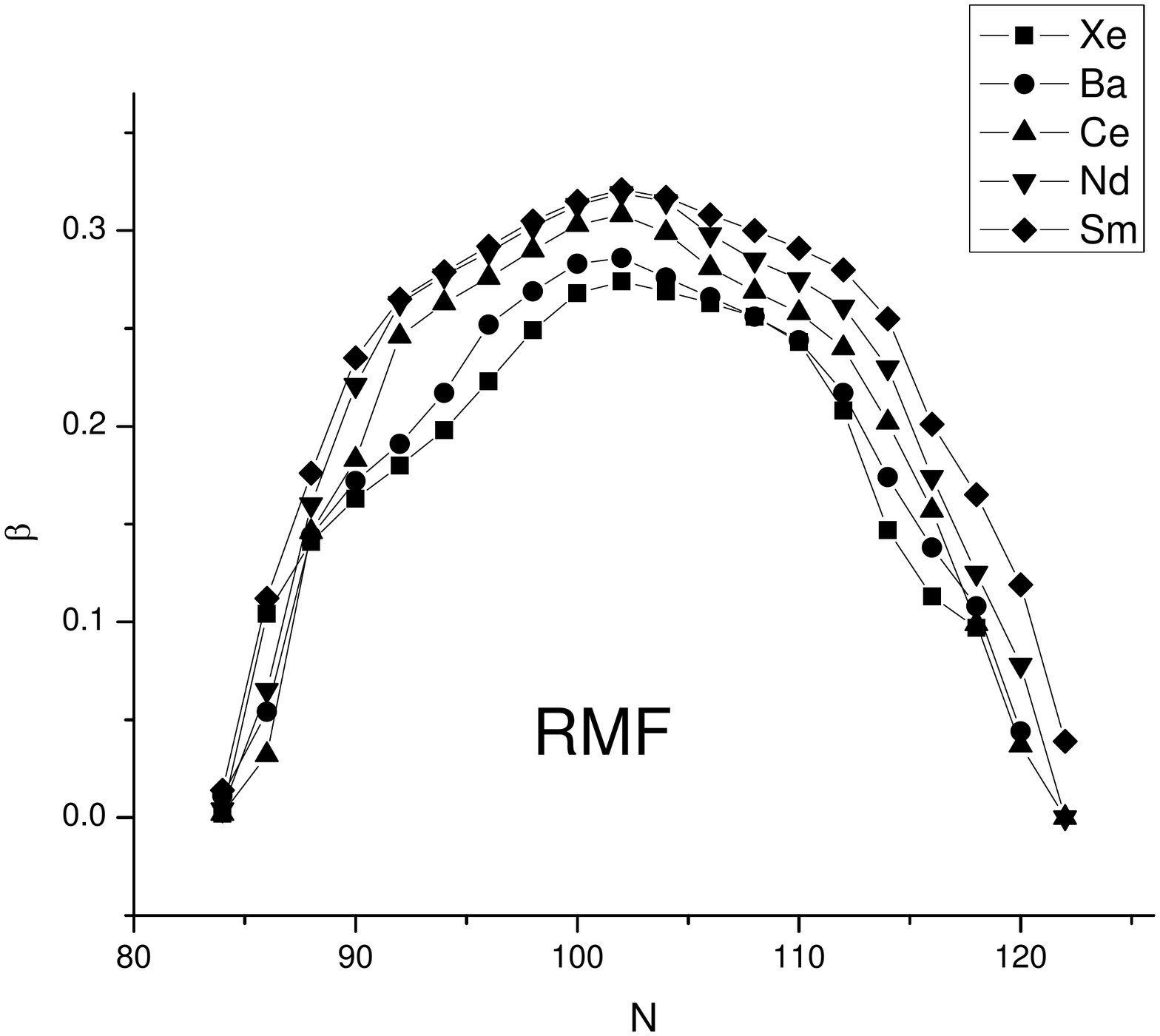}\hspace{5mm}
\includegraphics{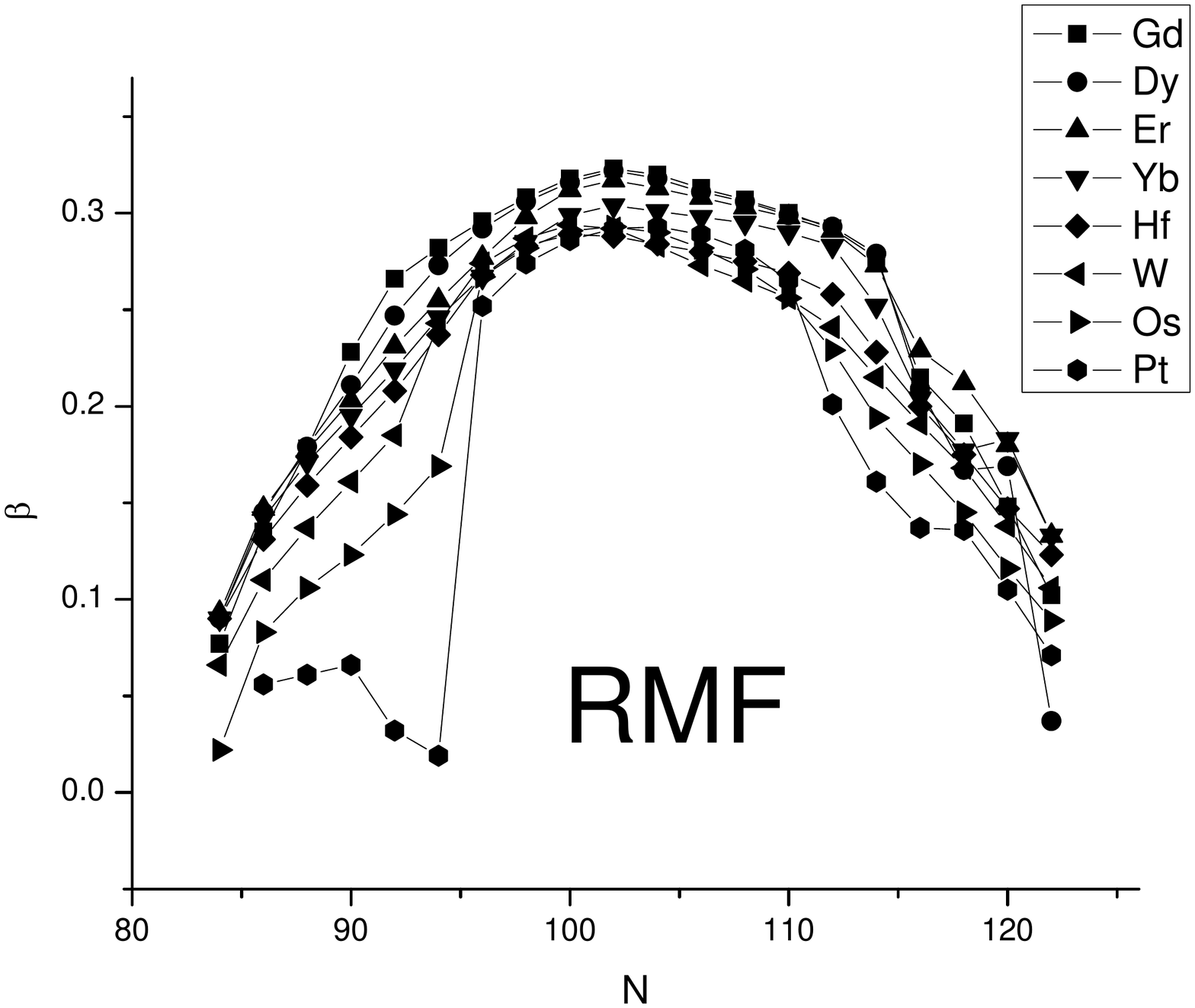}} 
}

\caption{RMF predictions for $\beta$, obtained from Ref. \cite{Lalazissis}. } 
\end{figure*}

\begin{figure*}[htb]

\resizebox{0.99\textwidth}{!}{%

{\includegraphics{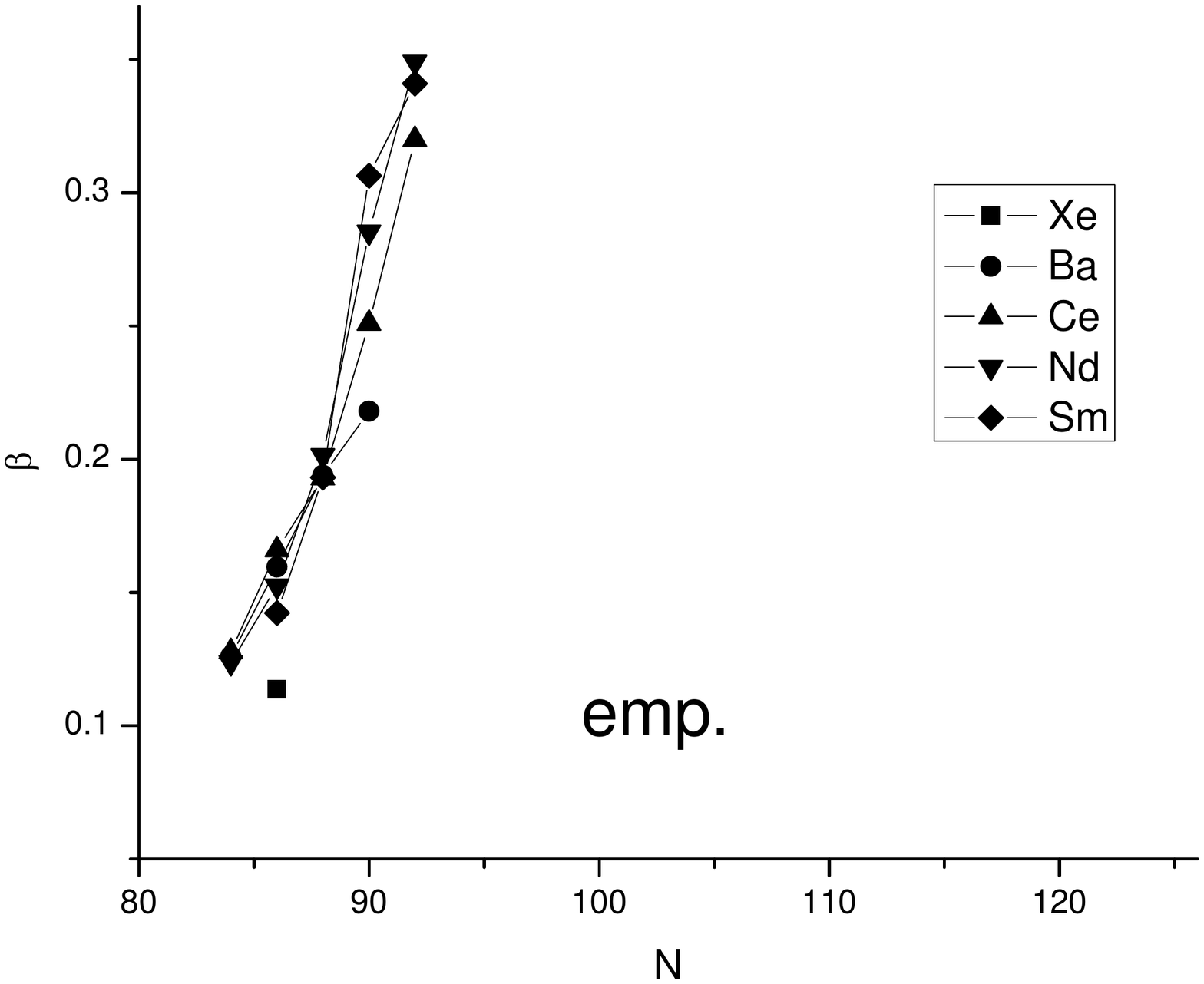}\hspace{5mm}
\includegraphics{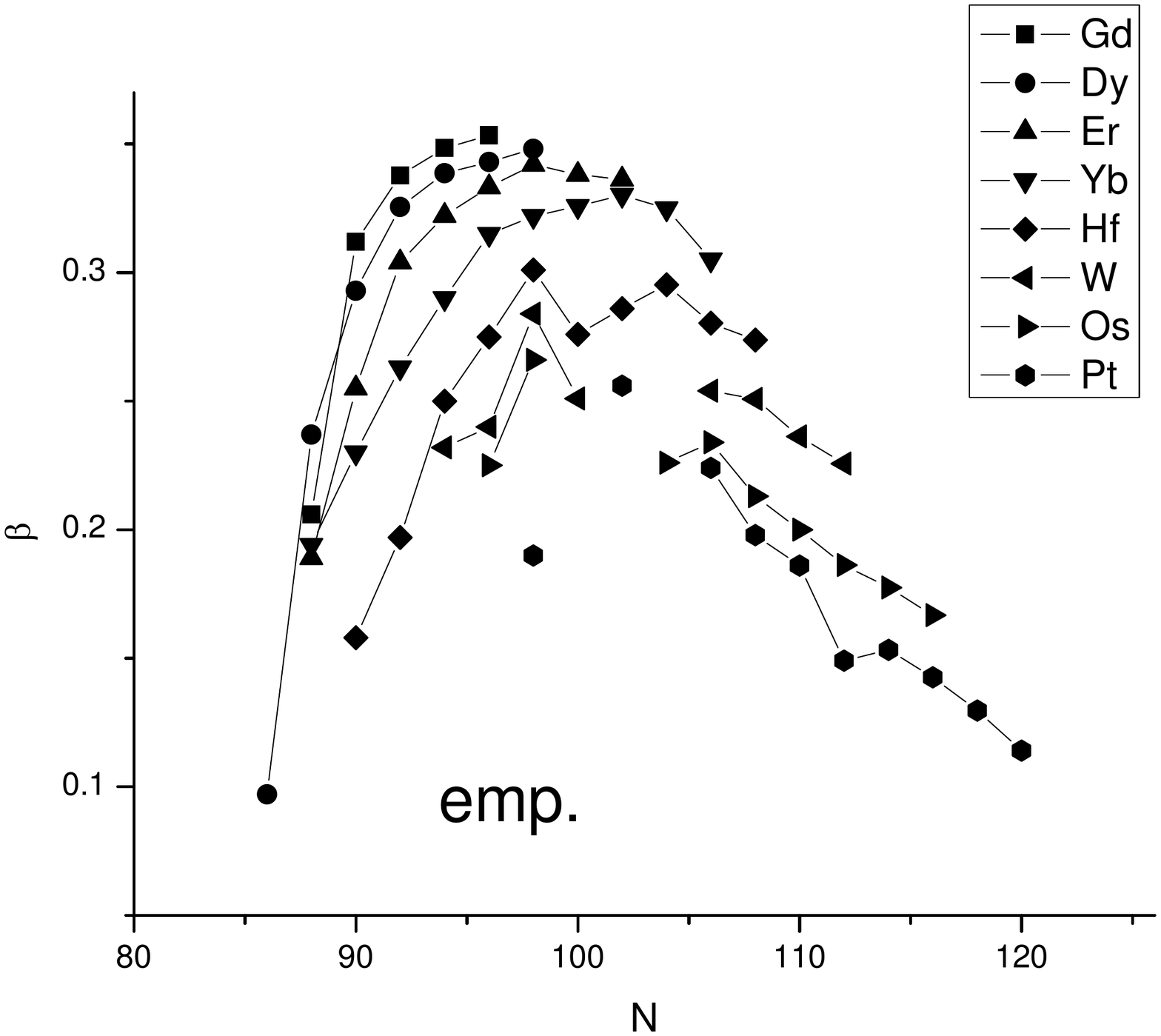}} 
}

\caption{Empirical predictions for $\beta$, obtained from Ref. \cite{Raman}. 
 } 
\end{figure*}

\begin{figure*}[htb]

\resizebox{0.90\textwidth}{!}{%

{\includegraphics{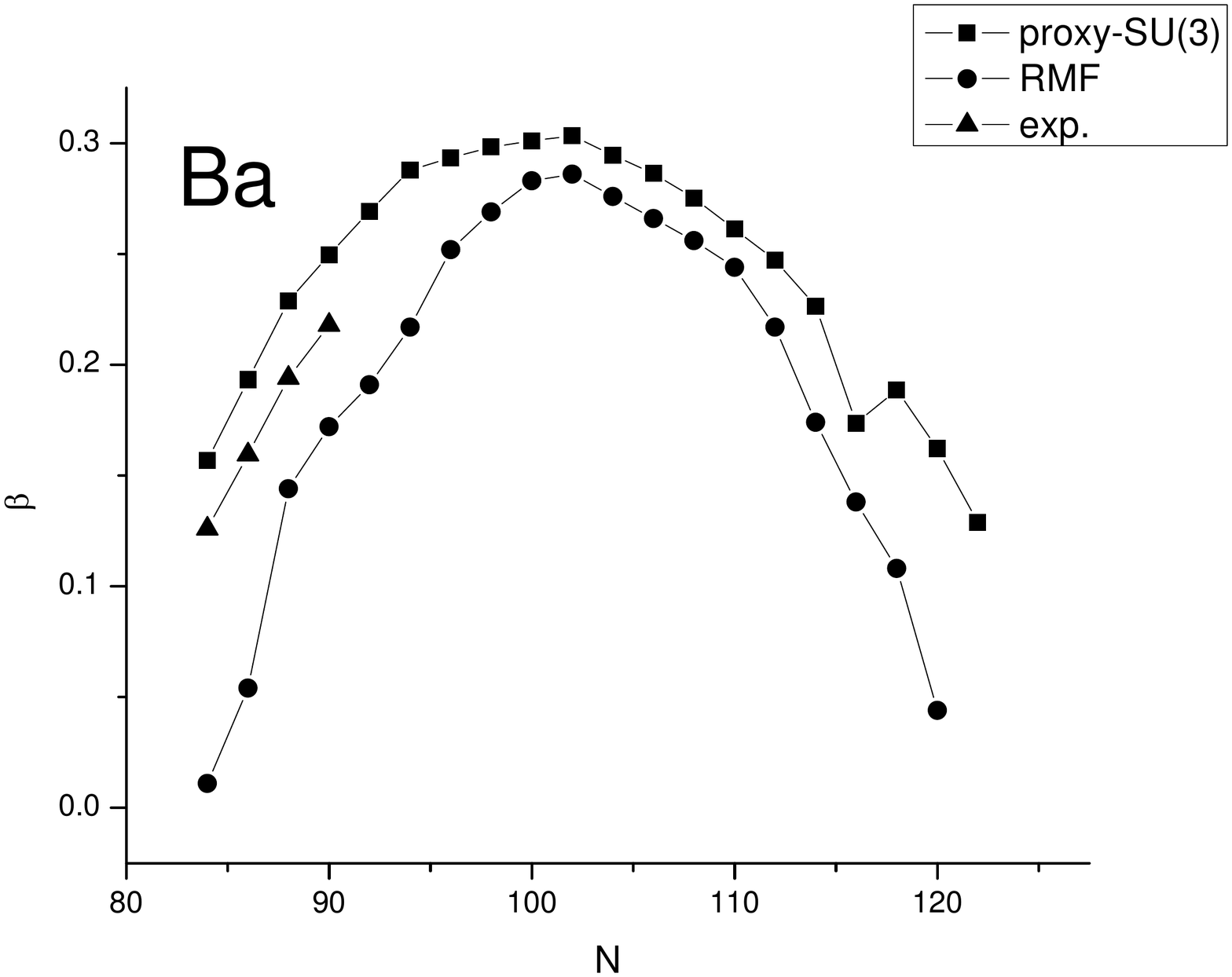}\hspace{5mm}
\includegraphics{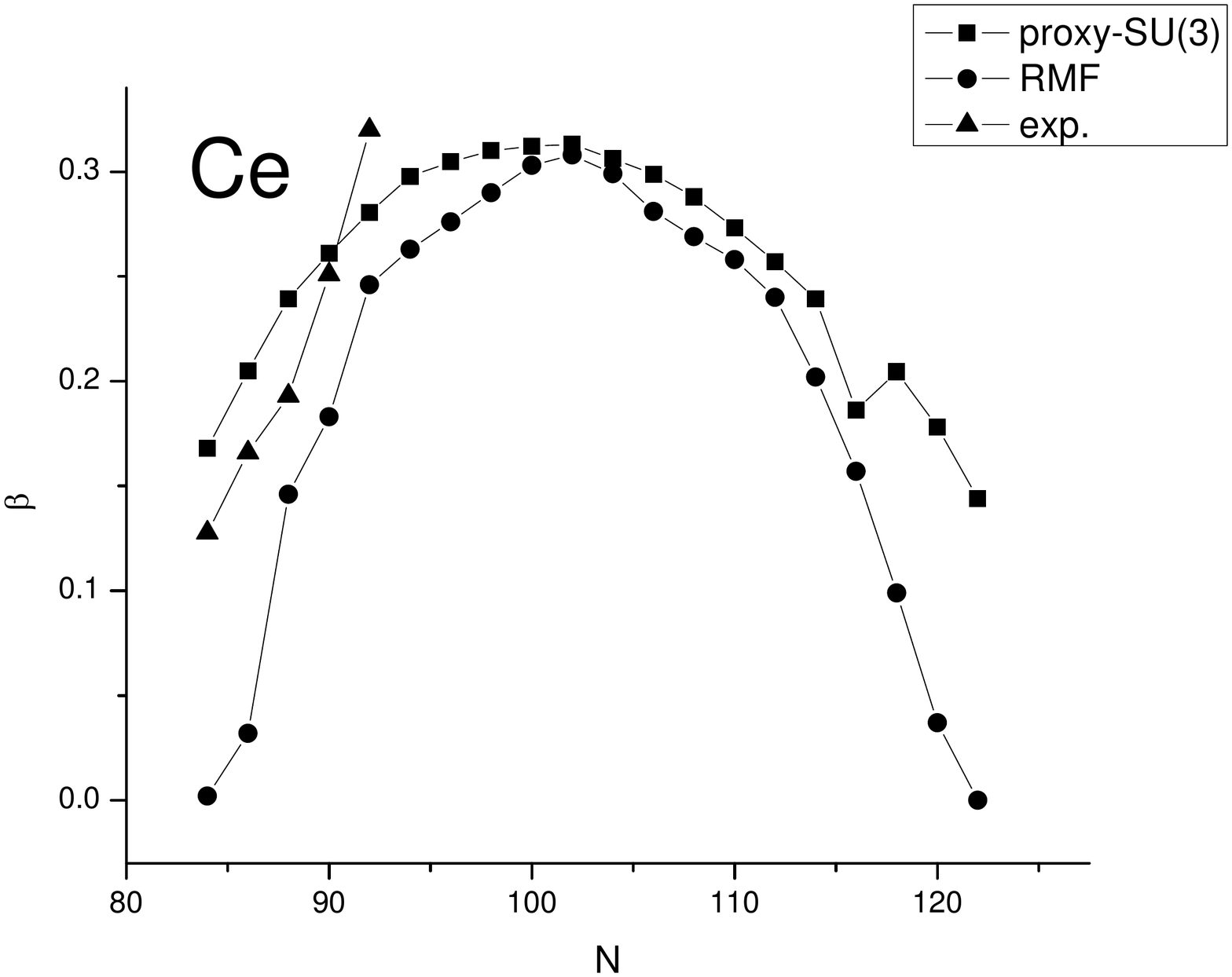}} 
}

\caption{Proxy-SU(3) predictions for $\beta$, obtained from Eq. (1), compared with tabulated $\beta$ values \cite{Raman} and also with predictions from relativistic mean field theory \cite{Lalazissis}. 
 } 
\end{figure*}

\begin{figure*}[htb]

\resizebox{0.90\textwidth}{!}{%

{\includegraphics{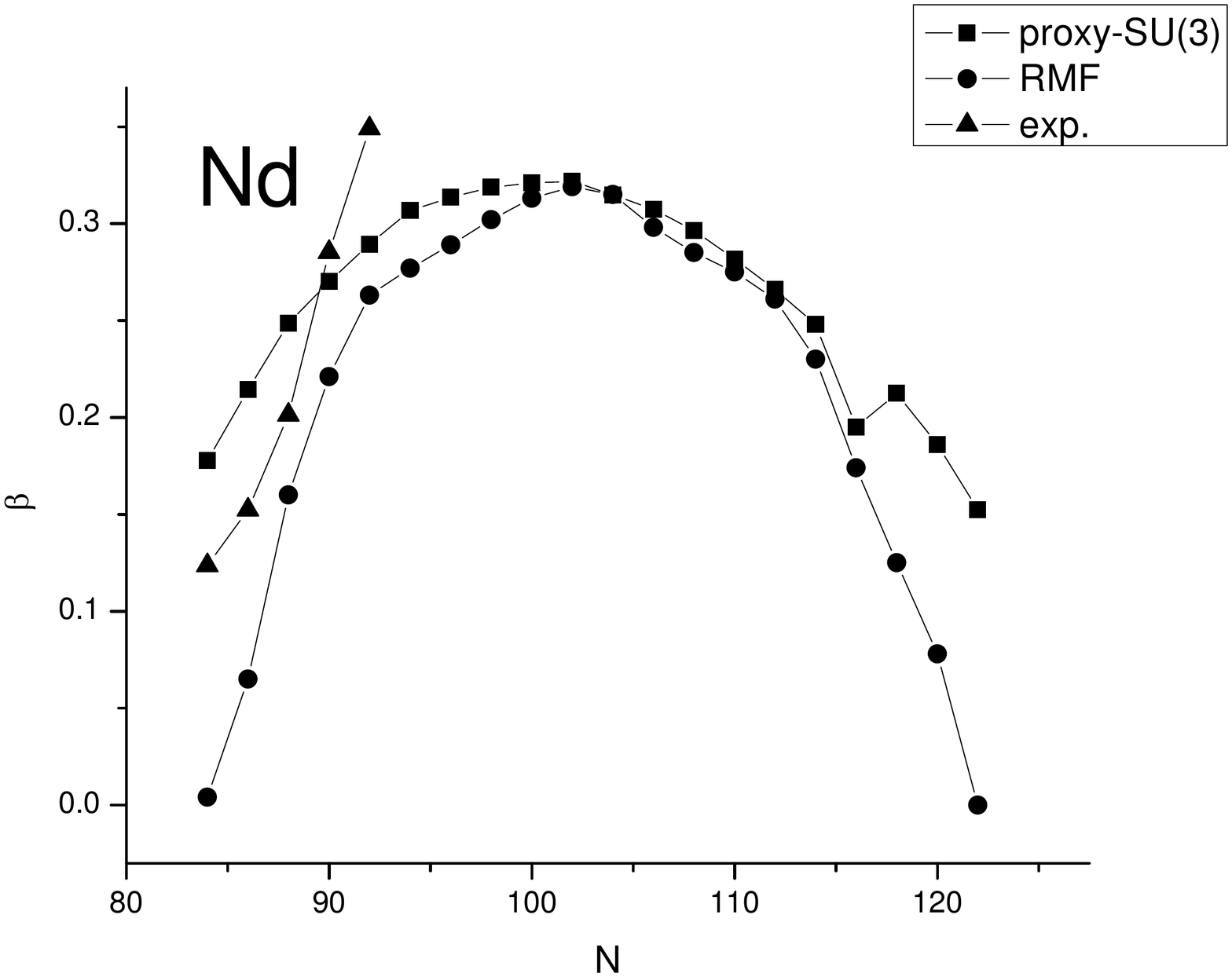}\hspace{5mm}
\includegraphics{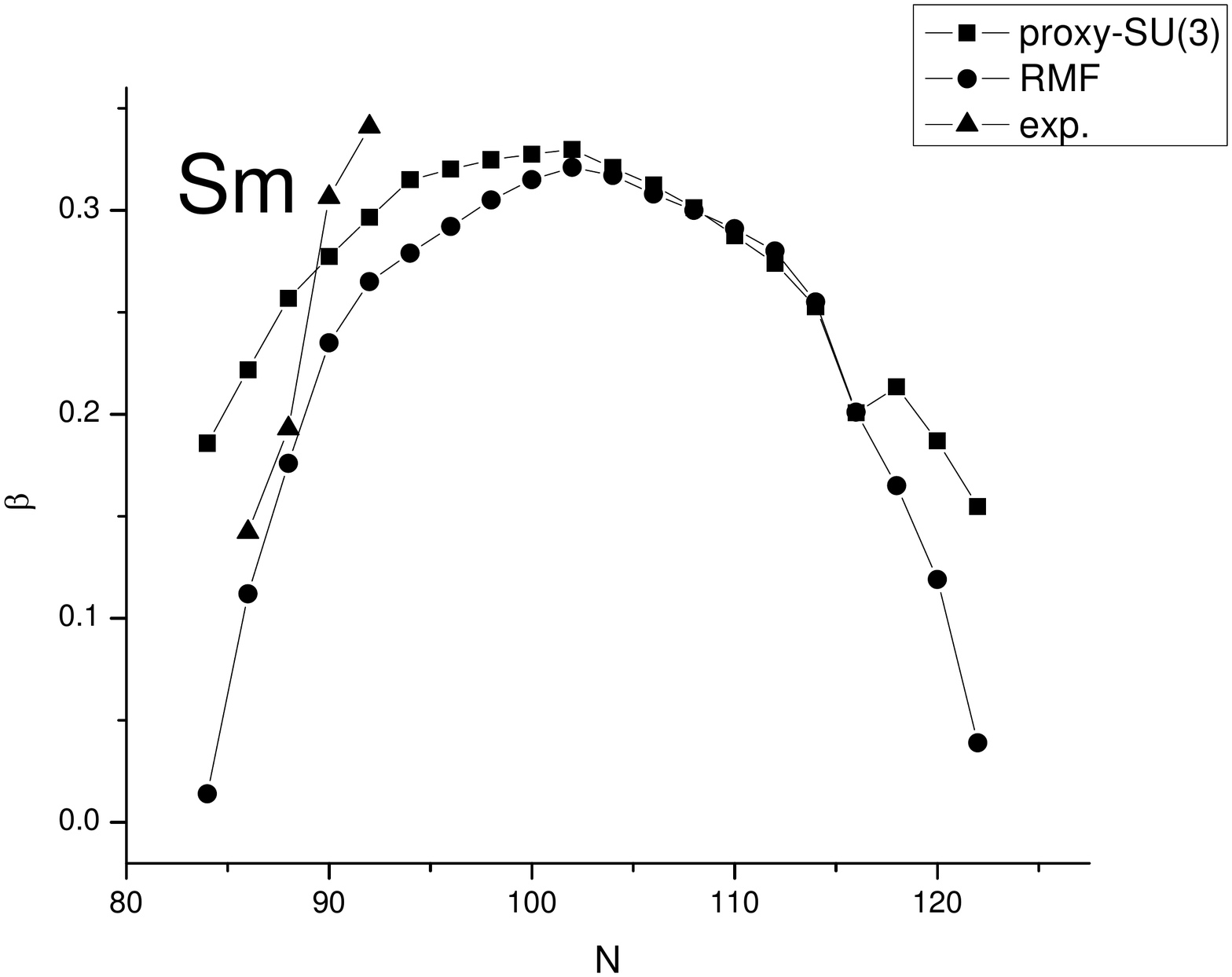}} 
}

\caption{Proxy-SU(3) predictions for $\beta$, obtained from Eq. (1), compared with tabulated $\beta$ values \cite{Raman} and also with predictions from relativistic mean field theory \cite{Lalazissis}. 
 } 
\end{figure*}

\begin{figure*}[htb]

\resizebox{0.90\textwidth}{!}{%

{\includegraphics{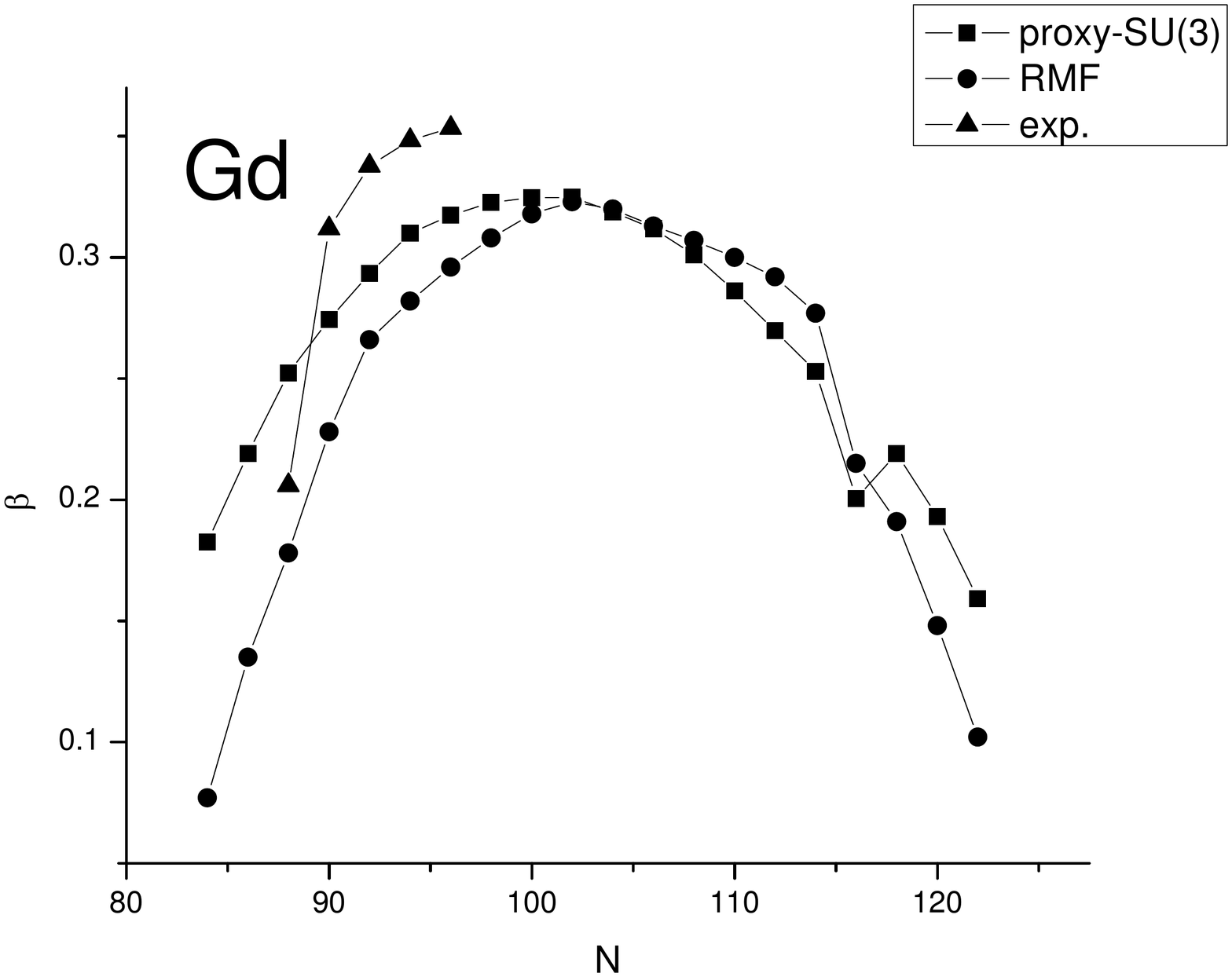}\hspace{5mm}
\includegraphics{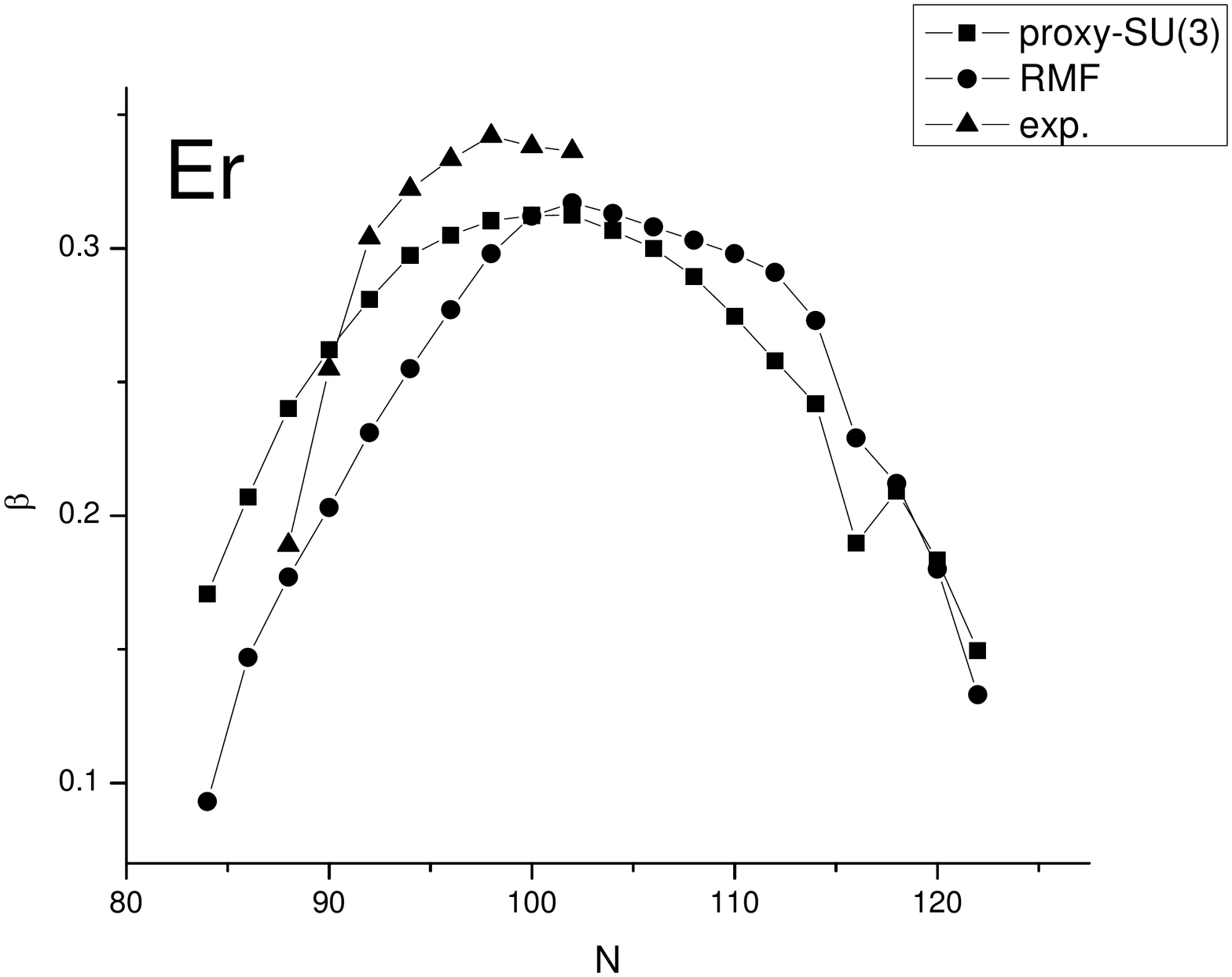}}
}

\caption{Proxy-SU(3) predictions for $\beta$, obtained from Eq. (1), compared with tabulated $\beta$ values \cite{Raman} and also with predictions from relativistic mean field theory \cite{Lalazissis}. 
 } 
\end{figure*}

\begin{figure*}[htb]

\resizebox{0.90\textwidth}{!}{%

{\includegraphics{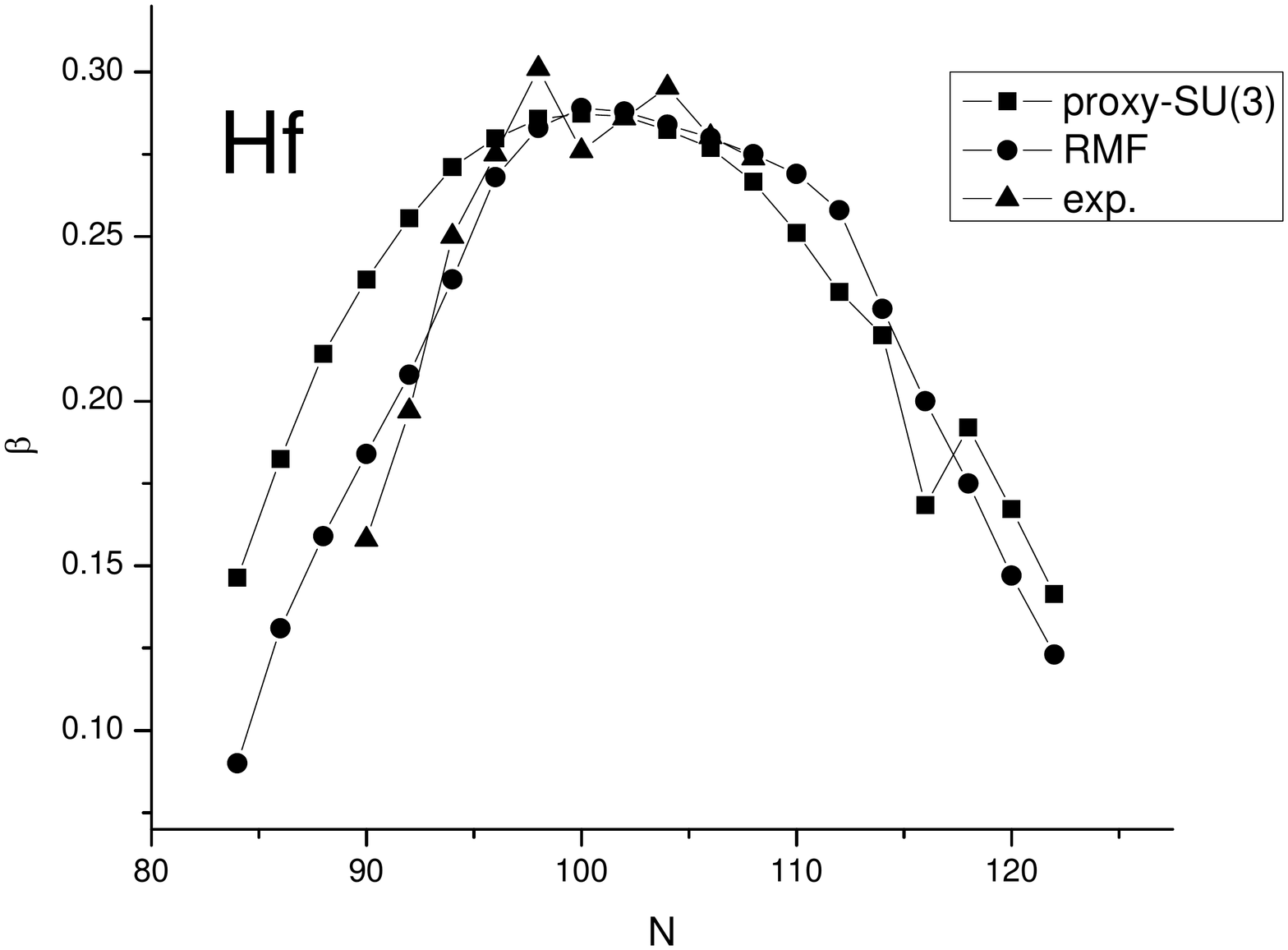}\hspace{5mm}
\includegraphics{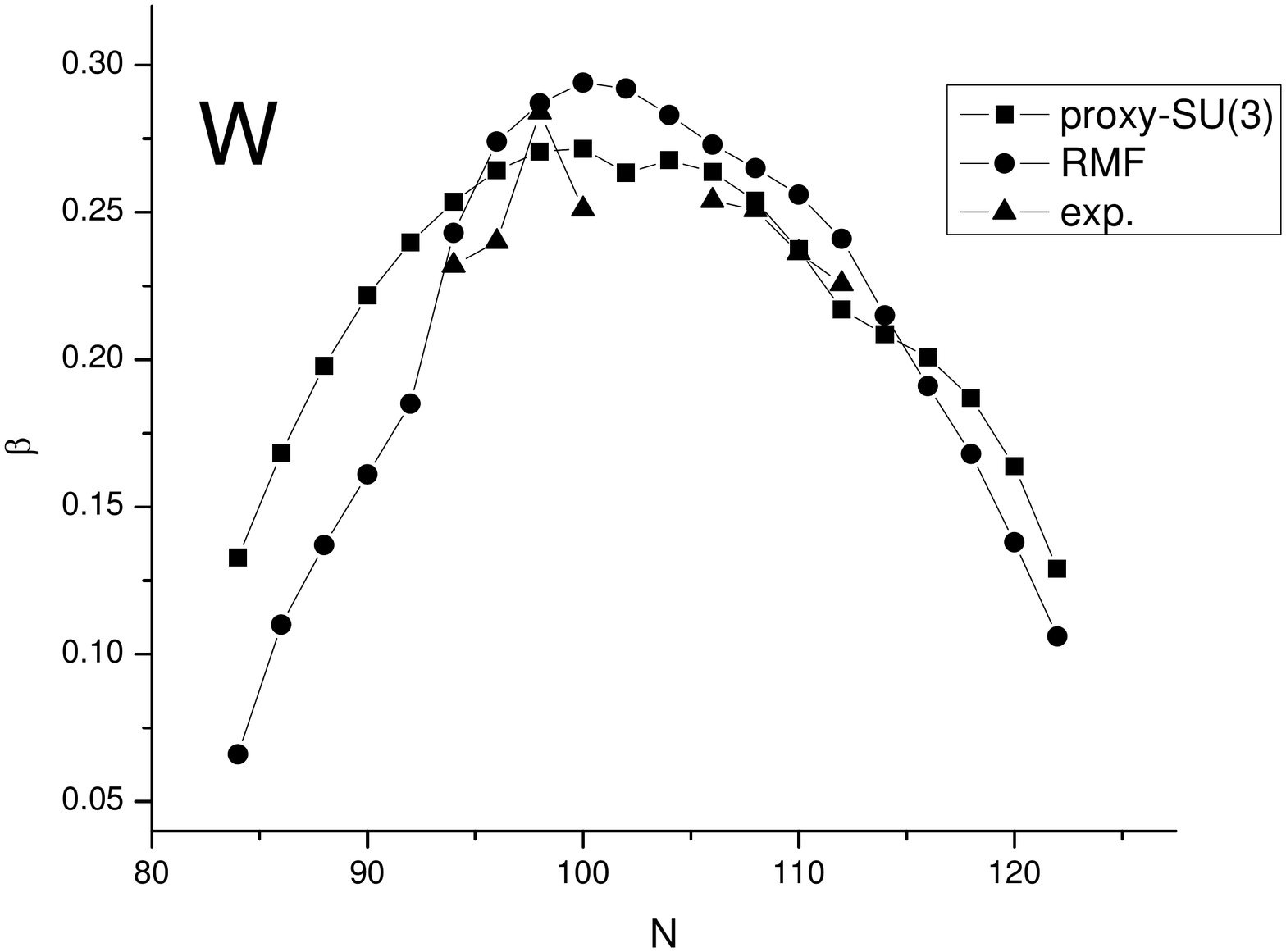}} 
}

\caption{Proxy-SU(3) predictions for $\beta$, obtained from Eq. (1), compared with tabulated $\beta$ values \cite{Raman} and also with predictions from relativistic mean field theory \cite{Lalazissis}. 
 } 
\end{figure*}

\begin{figure*}[htb]

\resizebox{0.90\textwidth}{!}{%

{\includegraphics{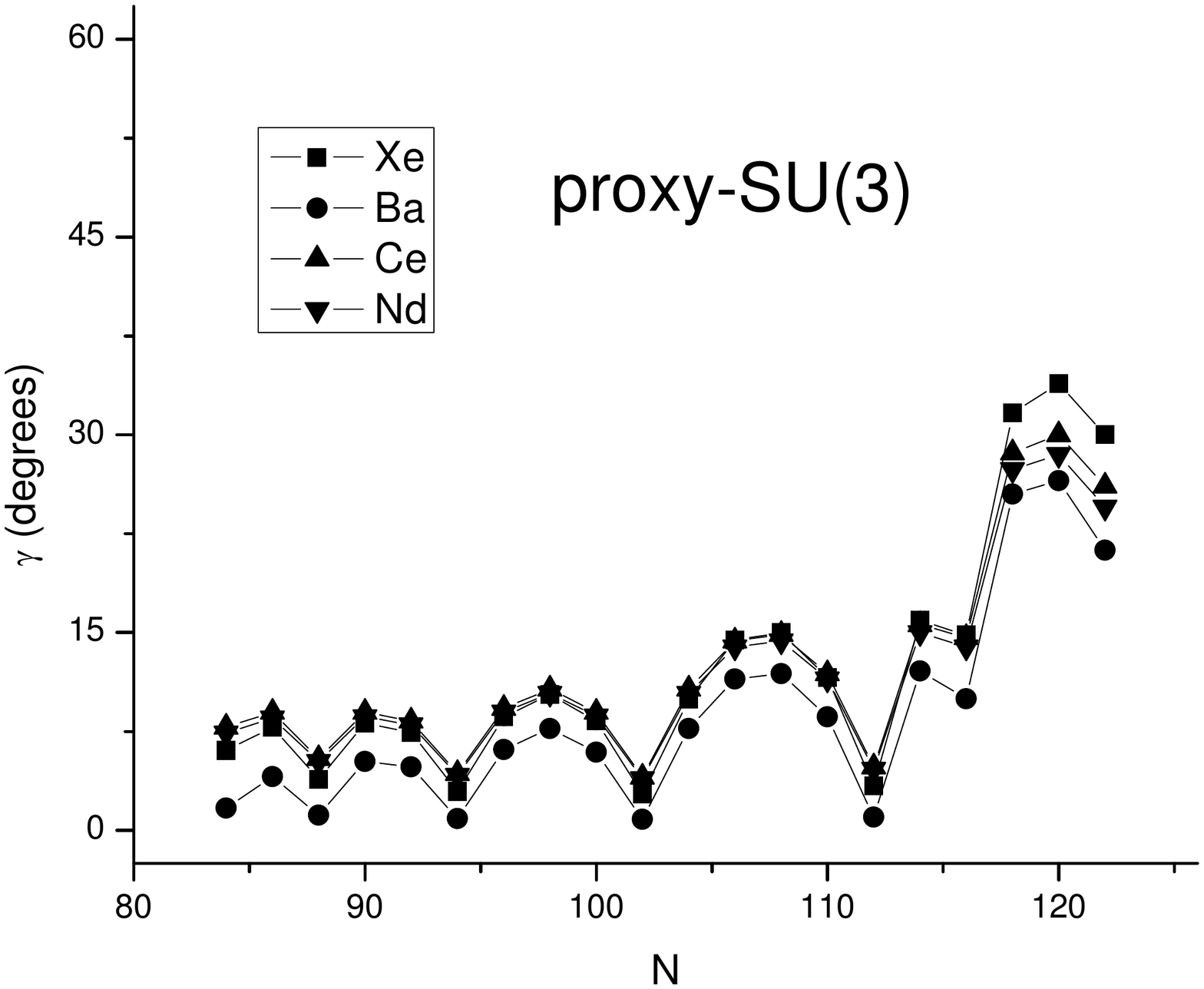}\hspace{5mm}
\includegraphics{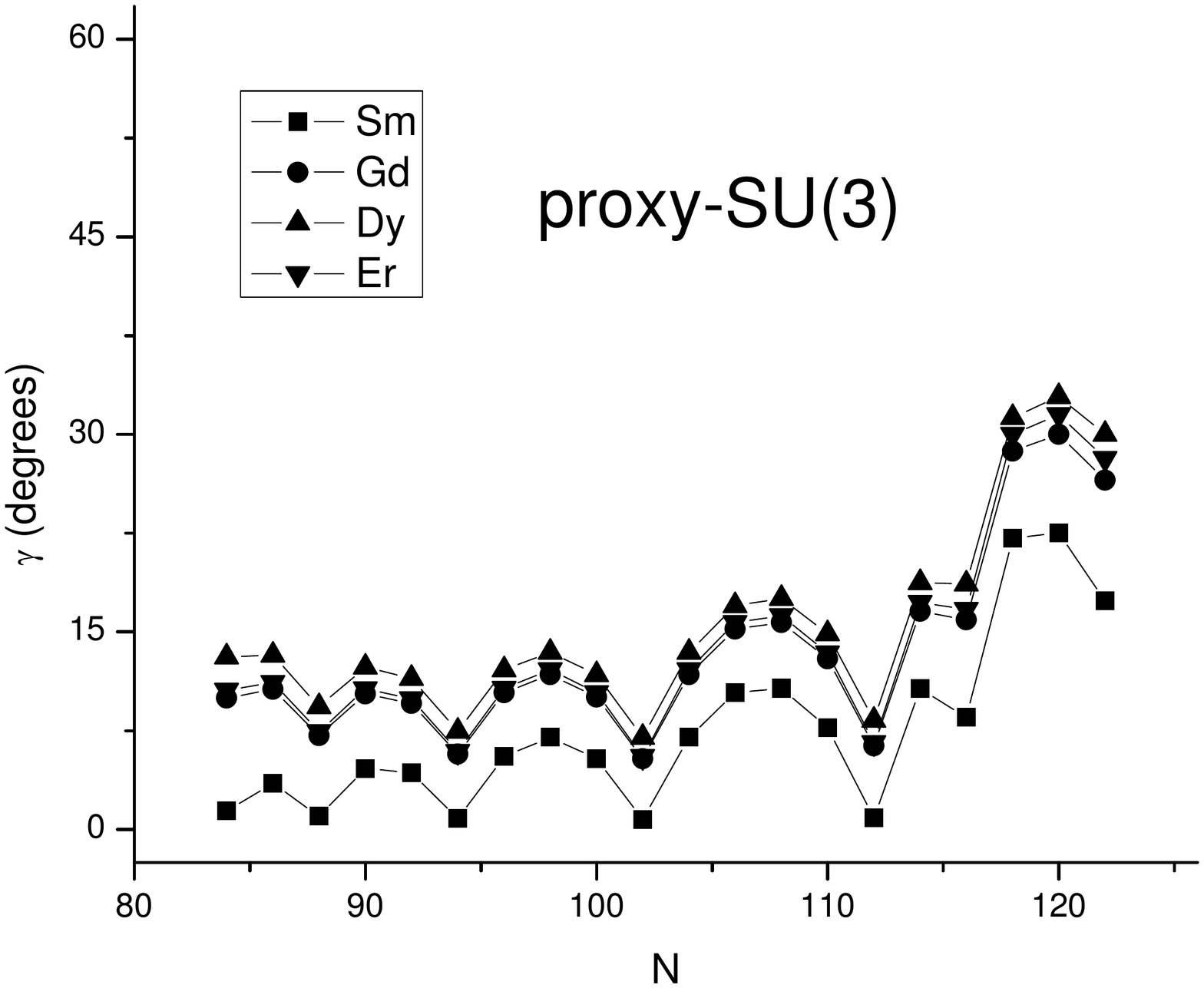}} 
}

\caption{Proxy-SU(3) predictions for $\gamma$, obtained from Eq. (3).
 } 
\end{figure*}

\begin{figure*}[htb]

\resizebox{0.90\textwidth}{!}{%

{\includegraphics{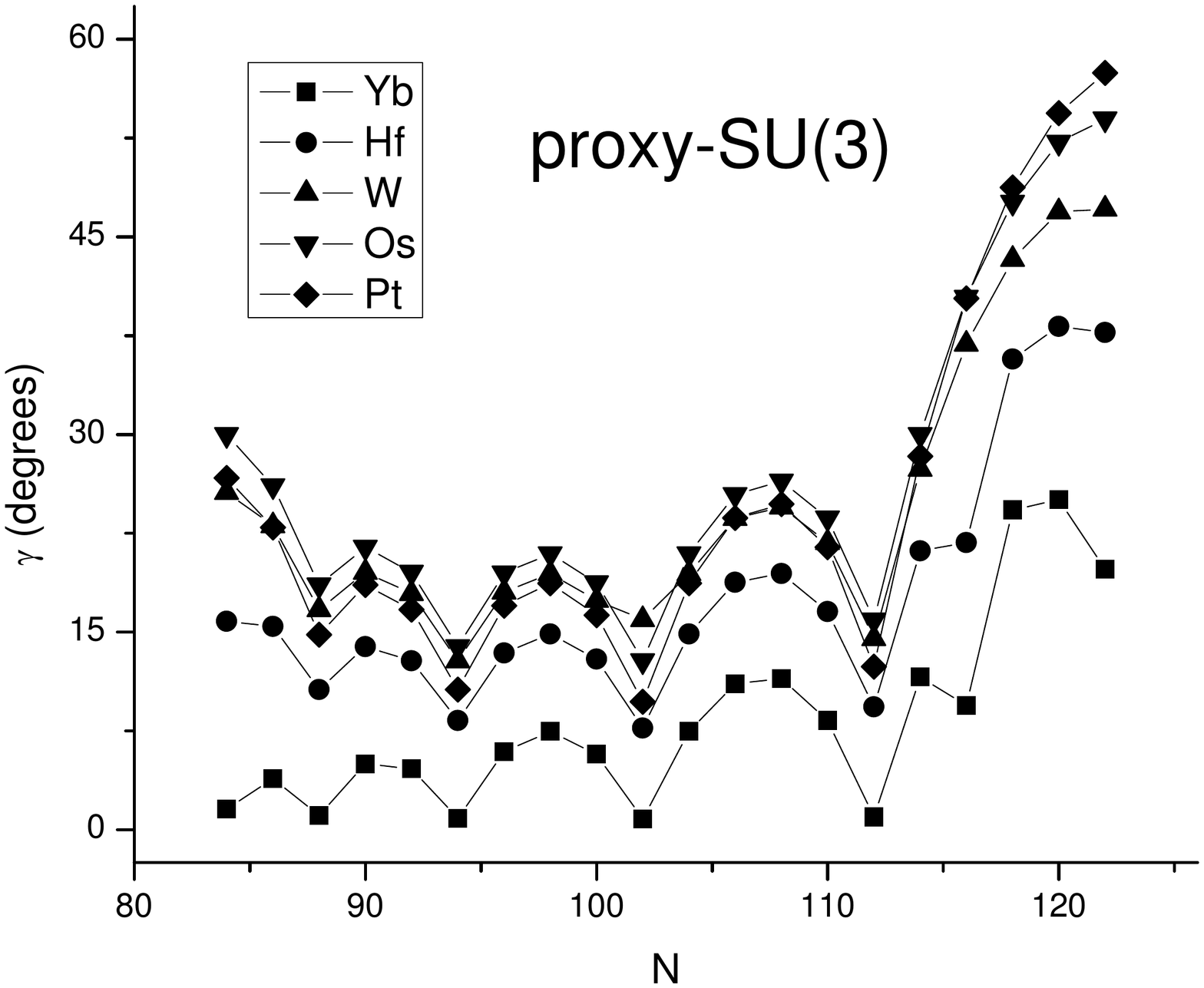}\hspace{5mm}
\includegraphics{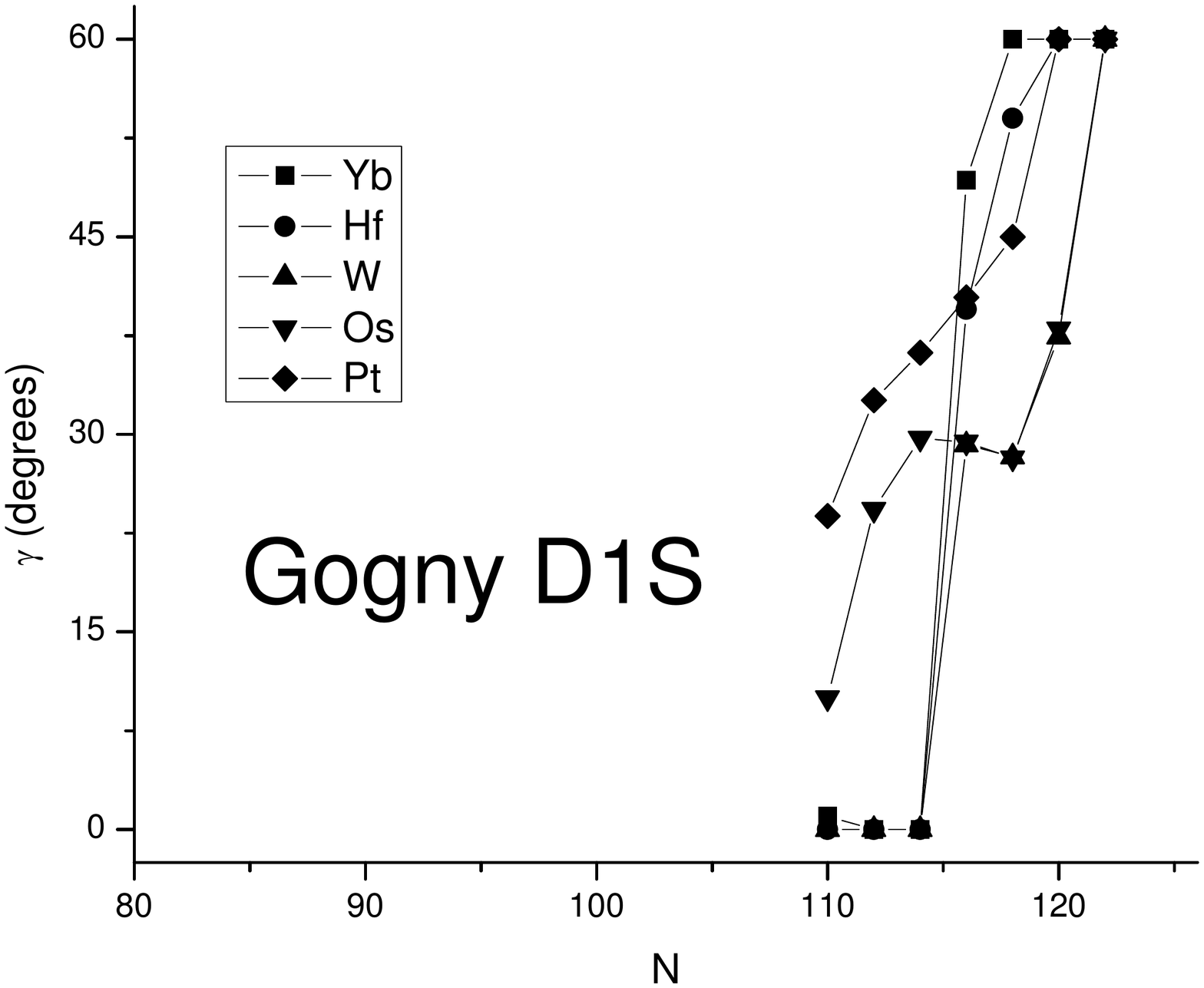}} 
}

\caption{Proxy-SU(3) predictions for $\gamma$, obtained from Eq. (3) and from Gogny D1S calculations \cite{Robledo}.
 } 
\end{figure*}

\begin{figure*}[htb]

\resizebox{0.90\textwidth}{!}{%

{\includegraphics{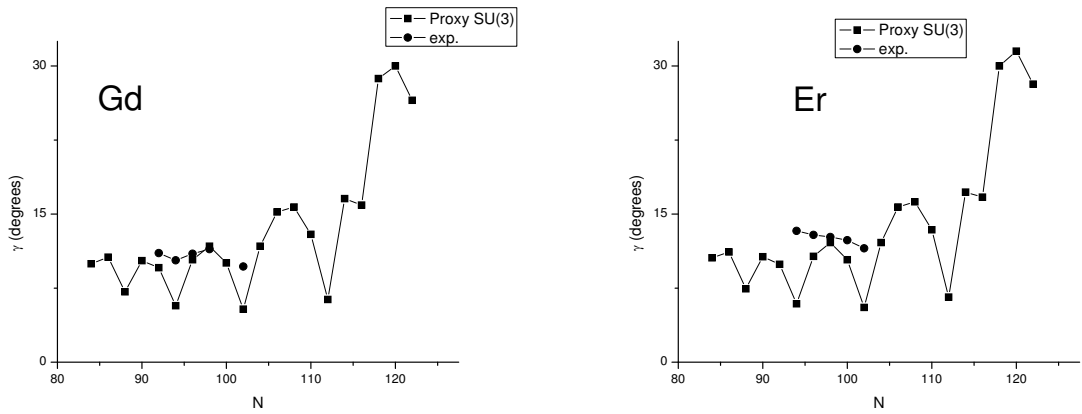}\hspace{5mm}
\includegraphics{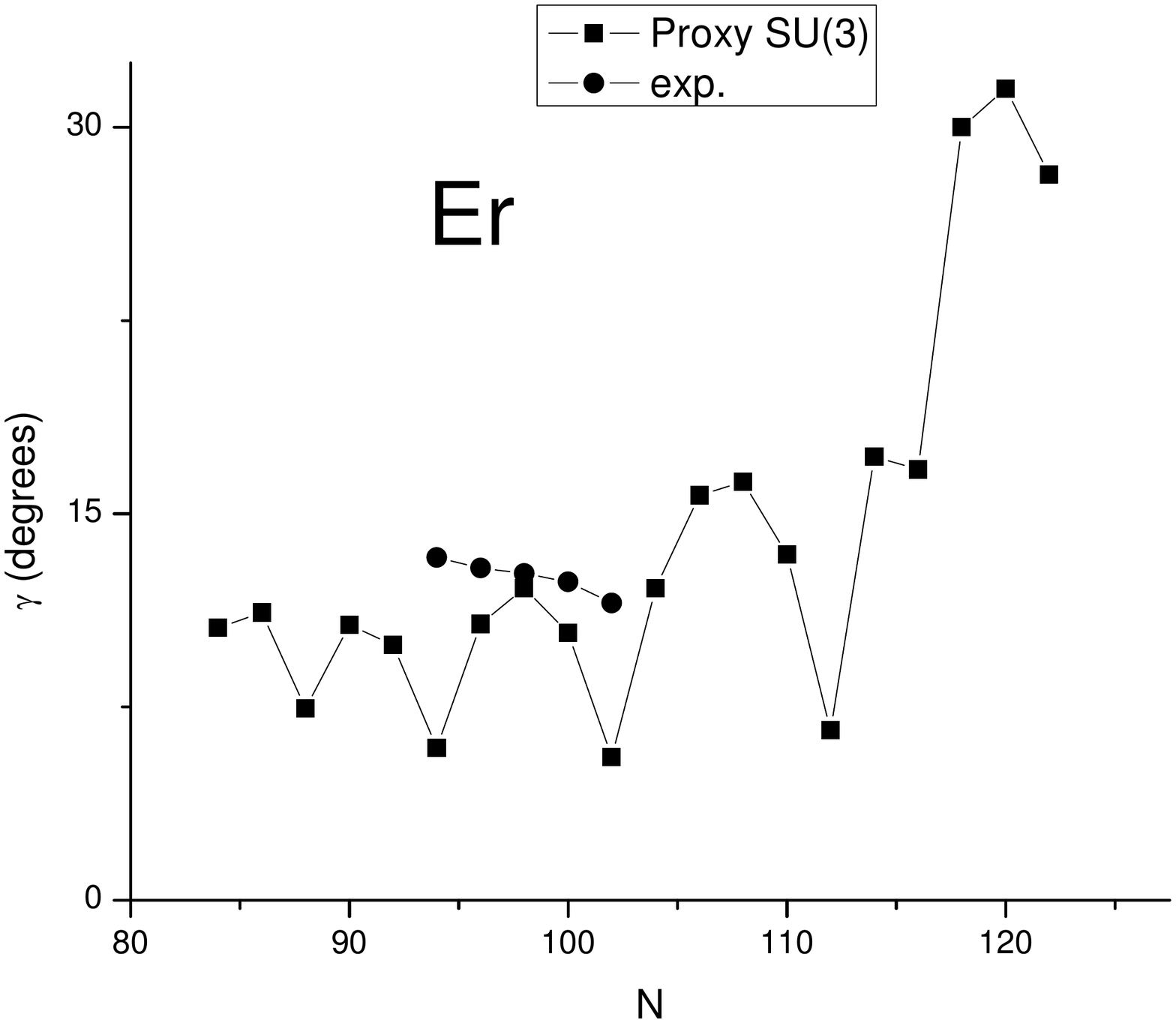}} 
}

\caption{Proxy-SU(3) predictions for $\gamma$, obtained from Eq. (3), compared with 
experimental values obtained from Eq. (5) \cite{Casten,Esser}, as well as with  predictions
of Gogny D1S calculations \cite{Robledo}.
 } 
\end{figure*}

\begin{figure*}[htb]

\resizebox{0.90\textwidth}{!}{%

{\includegraphics{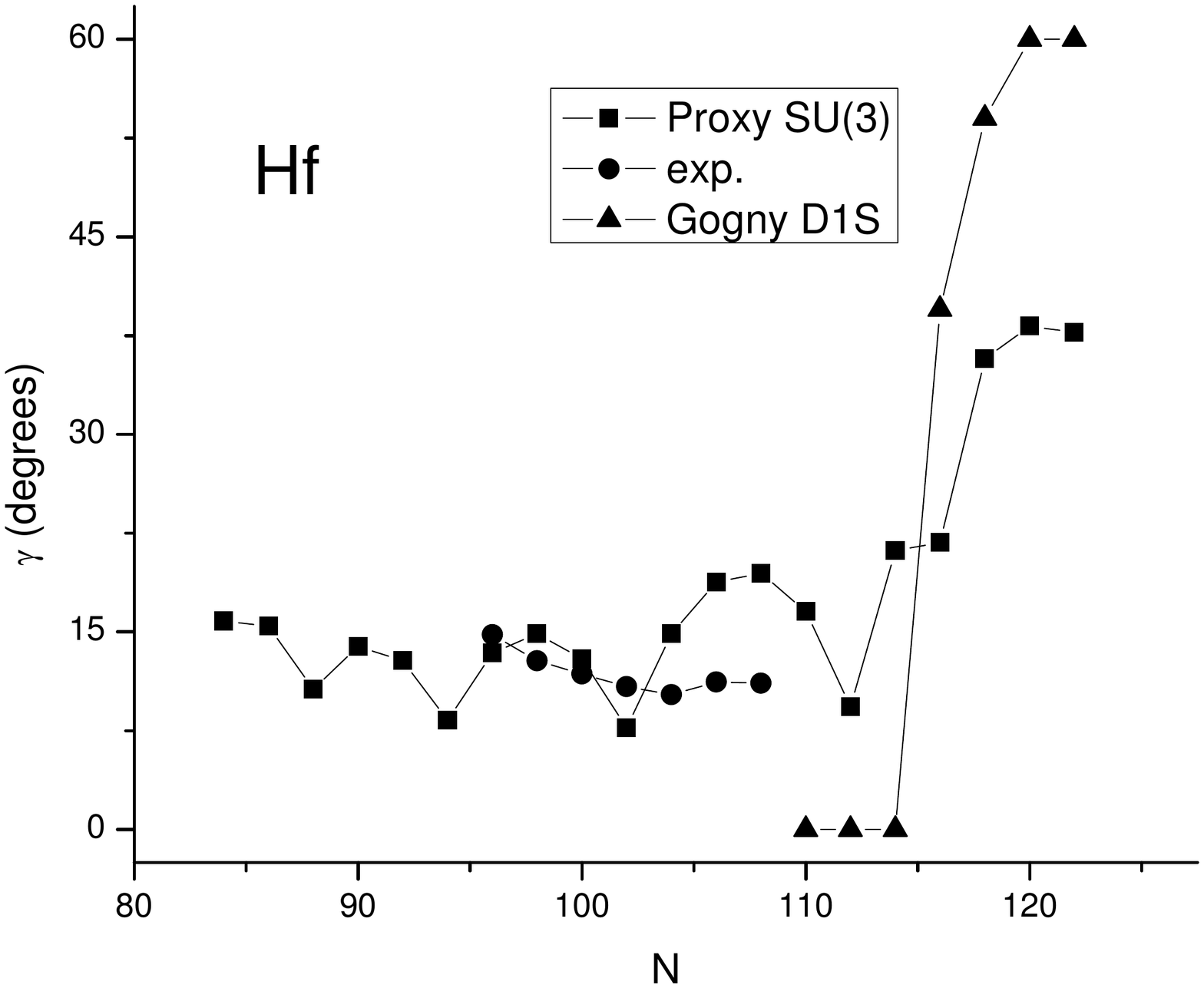}\hspace{5mm}
\includegraphics{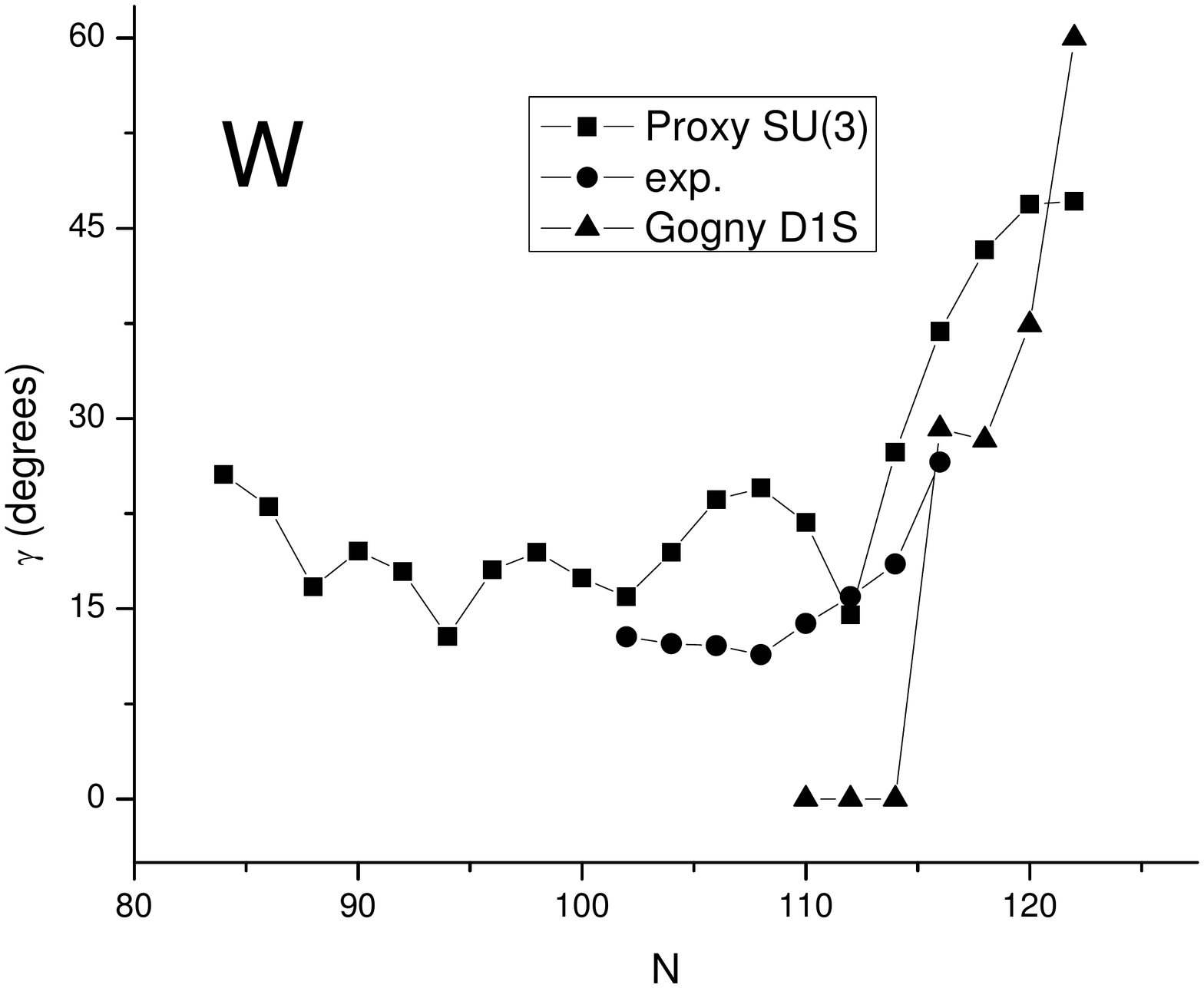}} 
}

\caption{Proxy-SU(3) predictions for $\gamma$, obtained from Eq. (3), compared with 
experimental values obtained from Eq. (5) \cite{Casten,Esser}, as well as with  predictions
of Gogny D1S calculations \cite{Robledo}.
 } 
\end{figure*}

\end{document}